\documentclass[12pt]{article}
\usepackage[width=\textwidth,small]{caption}
\usepackage[colorlinks=true]{hyperref}
\usepackage{fancyvrb}
\usepackage{graphicx}
\usepackage{upgreek}
\usepackage{amsmath}
\usepackage{amssymb}
\usepackage{times}
\usepackage{color}
\usepackage{bm}

\usepackage{caption}

\definecolor{red}{rgb}{0,0,0}
\definecolor{blue}{rgb}{0,0,0.75}
\definecolor{green}{rgb}{0,0.5,0}

\newcommand{\D}[0]{{\rm d}}

\begin{document}
\title{\bf Contour models of cellular adhesion}
\author{Luca Giomi \\ {\em\small Instituut-Lorentz, Universiteit Leiden, P.O. Box 9506, 2300 RA Leiden, The Netherlands}}
\date{}
\maketitle

\begin{abstract}
The development of traction-force microscopy, in the past two decades, has created the unprecedented opportunity of performing direct mechanical measurements on living cells as they adhere or crawl on unifrom or micro-patterend substrates. Simultaneously, this has created the demand for a theoretical framework able to decipher the experimental observations, shed light on the complex biomechanical processes that govern the interaction between the cell and the extracellular matrix and offer testable predictions. Contour models of cellular adhesion, represent one of the simplest and yet most insightful approach in this problem. Rooted in the paradigm of active matter, these models allow to explicitly determine the shape of the cell edge and calculate the traction forces experienced by the substrate, starting from the internal and peripheral contractile stresses as well as the passive restoring forces and bending moments arising within the actin cortex and the plasma membrane. In this chapter I provide a general overview of contour models of cellular adhesion and review the specific cases of cells equipped with isotropic and anisotropic actin cytoskeleton as well as the role of bending elasticity.\\[10pt]
\end{abstract}

In this chapter we review one of the simplest and yet most insightful theoretical approach for modeling the mechanics of cell-substrate interaction, based on two-dimensional contour models. Introduced by Bar-Ziv {\em et al}. to account for the pearling instability in fibroblasts \cite{BarZiv:1999} and later systematically developed by Bischofs {\em et al.} \cite{Bischofs:2008,Bischofs:2009} in the context of traction force microscopy, this approach consists of modeling cells adhering on uniform or micro-patterned  substrates as two-dimensional active gels subject to internal and external forces. This approach allows an explicit calculation of both the shape of the cell and the traction forces exerted by the cell on the substrate, under the assumption that the time scale required for the equilibration of the internal forces is much shorter than the typical time required by the cell to move across the substrate (i.e. minutes). The seemingly intractable problem of predicting the shape of a living cell is then brought into the realm of classical continuum mechanics, with the task of modeling the internal passive and active forces representing the major technical challenge. Starting from the original work by Bar-Ziv {\em et al}. \cite{BarZiv:1999}, much progress has been made on incorporating into this simple picture aspects of the mechanical complexity of eukaryotic cells, including the bending elasticity of the plasma membrane \cite{Banerjee:2013} and the anisotropy of the actin cytoskeleton \cite{Pomp:2018}. Whether the final product of this reductionist effort is still far from providing a theoretical framework whose predictive power is comparable with the most recent computational approaches \cite{Sabass:2008}, it nevertheless represents an indispensable conceptual step toward a satisfactory understanding of the physical properties of the cell.

From simple prokaryotes to the more complex eukaryotes, living cells are capable of astonishing mechanical functionalities \cite{Jamney:2007}. They can repair wounded tissues by locally contracting the extracellular matrix \cite{Midwood:2004}, move in a fluid or on a substrate \cite{Barry:2010}, and generate enough force to split themselves in two while remaining alive \cite{Tanimoto:2012}. Conversely, cell behavior and fate crucially depend on mechanical cues from outside the cell \cite{Geiger:2009,Discher:2005,Julicher:2007,Mendez:2012}. Examples include rigidity-dependent stem cell differentiation \cite{Engler:2006,Trappmann:2012}, protein expression regulated by internal stresses \cite{Sawada:2006}, mechanical cell-cell communication \cite{Reinhart-King:2008} and durotaxis \cite{Lo:2000,Sochol:2011}. In all these bio-mechanical processes, cells rely on their shape to gauge the mechanical properties of their microenvironment \cite{Ghibaudo:2009} and direct the traction forces exerted on their surrounding \cite{Schwarz:2013}.

\begin{figure}[t]
\centering
\includegraphics[width=0.7\textwidth]{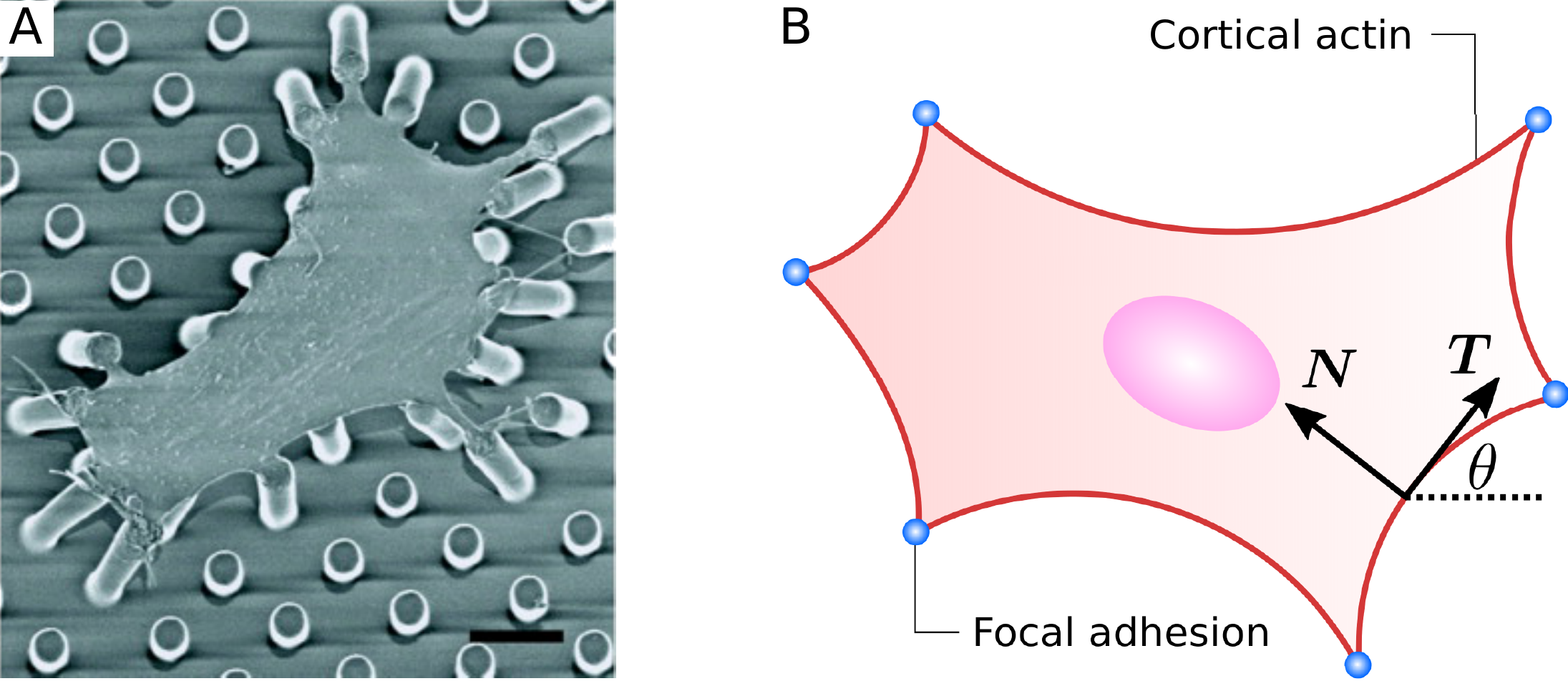}
\caption{\label{fig:schematic}(A) Scanning electron micrograph of a smooth muscle cell attached to an array of posts coated at the tip with fibronectin. Scale bar 10 $\upmu$m. Adapted from Ref. \cite{Tan:2003}. (B) Schematic representation of a contour model of adherent cells. The cell edge is parameterized as closed plane curve comprising concave arcs connecting pairs of adhesion points (blue dots). The geometry of the arcs can be entirely described via the two-dimensional Frenet-Serret frame consisting of the tangent vector $\bm{T}=(\cos\theta,\sin\theta)$ and the normal vector $\bm{N}=(-\sin\theta,\cos\theta)$, with $\theta$ the turning angle.}
\end{figure}

Immediately after coming into contact with such a surface, many animal cells spread and develop transmembrane adhesion receptors. This induces the actin cytoskeleton to reorganize into cross-linked networks and bundles \cite{Burridge:1996,Burridge:2013}, whereas adhesion becomes limited to a number of sites, distributed mainly along the cell contour (i.e. focal adhesions \cite{Burridge:1996}). At this stage, cells are essentially flat and assume a typical shape characterized by arcs which span between the sites of adhesion (Fig. 1a), while forces are mainly contractile \cite{Schwarz:2013}. On timescales much shorter than those required to change its shape, the cell can be considered in mechanical equilibrium at any point of its interface. Upon treating the cell as a two-dimensional continuum separated by the surrounding environment by a one-dimensional interface, such an equilibrium condition translates into the following force balance relation at the cell cortex: 
\begin{equation}\label{eq:force_balance}
\bm{F}'+(\bm{\hat{\Sigma}}_{\rm out}-\bm{\hat{\Sigma}}_{\rm in})\cdot\bm{N} + \bm{f}_{\rm ext}= \bm{0}\;,
\end{equation}
where $\bm{F}$ is the stress resultant along the cell cortex, the prime indicates differentiation with respect to the arc-length of the cell edge (i.e. $\bm{F}'=\D\bm{F}/\D s$), $\bm{\hat\Sigma}_{\rm out}$ and $\bm{\hat\Sigma}_{\rm in}$ are the stress tensors outside and inside the cell and $\bm{f}_{\rm ext}$ is an external force per unit length, possibly resulting from a deformation of the substrate. The cell contour is parametrized as a plane curve spanned by the arc-length $s$ and oriented along the inward pointing normal vector $\bm{N}$ (Fig. 1b). 

Eq. \eqref{eq:force_balance} is the starting point of all contour models of cellular adhesion as well as the core of this chapter. Before embarking on analyzing specific cases it is worth to stress that Eq. \eqref{eq:force_balance} is rooted into the following two fundamental assumptions: {\em 1)} cellular motion occurs {\em adiabatically}, because to the large separation of time scales associated with force relaxation and cell migration; {\em 2)} as a consequence of the flat morphology of adherent cells, one can ignore out-of-plane forces and reduce the dimension of the problem from three to two. 

In the remaining of this chapter, we will see how different bio-mechanical scenarios result in different choices of the forces and stresses involved in Eq. \eqref{eq:force_balance} and how these affect the shape of the cells and the traction forces experienced by the substrate. The chapter is organized as follows: in Sec. \ref{sec:math} we review some simple concept on plane curves and we fix a notation; in Sec. \ref{sec:stm} we introduce Bischofs' {\em et al}. simple tension model for isotropic contractile cells; in Sec. \ref{sec:atm} we discuss the case of anisotropic cells in the framework of the anisotropic tension model; in Sec. \ref{sec:bending} we explore the effect of bending elasticity; Sec. \ref{sec:conclusions} is devoted to conclusive remarks.

\section{\label{sec:math}Mathematical preliminaries and notation}

Whereas different in the mechanical details, all contour models of cellular adhesion revolve around modeling cells as flat two-dimensional objects whose edge can be mathematically described as a closed, but not necessarily smooth, plane curve. The geometry of plane curves can be entirely described via a two-dimensional version of the Frenet-Serret frame, consisting of the tangent vector $\bm{T}$ and the normal vector $\bm{N}$ (Fig. \ref{fig:schematic}b). Letting $\bm{r}=\bm{r}(s)$ the position vector of a curve in $\mathbb{R}^{2}$, parametrized via the arc-length $s$, these are defined by the following identities:
\begin{equation}\label{eq:frenet}
\bm{r}' = \bm{T}\;,\qquad\qquad
\bm{T}' = \kappa\bm{N}\;,\qquad\qquad
\bm{N}' =-\kappa\bm{T}\;,
\end{equation}
where $\kappa=\kappa(s)$ is the local curvature of the curve. Eqs. \eqref{eq:frenet} describe how the orthonormal frame $\{\bm{T},\bm{N}\}$ rotates on the plane as we move along the curve. Once $\kappa$ is assigned, in the form of a differentiable function, the fundamental theorem of plane curves guarantees that the corresponding curve is uniquely determined, up to a rigid motion \cite{DoCarmo:1976}. Furthermore, the tangent vector can be parametrized through a single scalar function $\theta=\theta(s)$, representing the turning angle of $\bm{T}$. In a standard Cartesian frame: $\bm{T}=(\cos\theta,\sin\theta)$, then, using Eq. \eqref{eq:frenet}, one finds $\kappa=\D\theta/\D s$ and:
\begin{equation}
\bm{r}(s) = \bm{r}(0)+\int_{0}^{s}\D s'\,\left[\cos\theta(s')\,\bm{\hat{x}}+\sin\theta(s')\,\bm{\hat{y}}\right]\;,
\end{equation}
with $\theta(s)=\theta(0)+\int_{0}^{s}\D s\,\kappa(s')$. Expressing the curvature of a plane curve as the derivative of the turning angle $\theta$, allows us to unambiguously identify the sign of $\kappa$ (unlike in three dimensions). Thus $\kappa>0$ ($\kappa<0$) implies that the turning angle increases (decreases) as we move along the curve, whereas reversing the orientation of the curve changes the sign of $\kappa$. If a closed plane curve is oriented counterclockwise and the turning angle is measured, as usual, with respect to the $x-$axis, then $\kappa>0$ ($\kappa<0$) will then corresponds to points where the curve is convex (concave). 

Closed curves, have a number of interesting global properties that serve as important calculation tools in the later sections. The four vertex theorem \cite{DeTurck:2007}, is one of the earliest results in global differential geometry and states that the curvature of a simple, smooth, closed curve on a plane has at least four vertices: i.e. four extrema where $\D\kappa/\D s=0$ (specifically two maxima and two minima). Another fundamental property of closed plane curves is expressed by the theorem of turning tangents \cite{DoCarmo:1976,Gray:1997}. Namely:
\begin{equation}\label{eq:turning_tangents}
\oint {\rm d}s\,\kappa = 2\pi m\;.	
\end{equation} 
The integer $m$ is called the rotation index of the curve and measures how many times the tangent vector turns with respect to a fixed direction \cite{Gray:1997}. Simple closed curves have thus $m=1$, whereas a curve that loops twice around its center (thus self-intersects once before closing) has $m=2$. If a simple closed curve has kinks (i.e. singular points where the tangent vector switches discontinuously between two orientations), these will affect the total curvature as follows:
\begin{equation}\label{eq:kinks}
\oint {\rm d}s\,\kappa + \sum_{i}\vartheta_{i} = 2\pi m\;.	
\end{equation} 
where $\vartheta_{i}$ is the external angle at each kink and the summation runs over all the kinks. In the case of a convex polygon, for instance, $\kappa=0$ and Eq. \eqref{eq:kinks} asserts that the sum of the external angles is equal to $2\pi$. 

Finally, by virtue of the divergence theorem in two dimensions, the area enclosed by a plane curve can be expressed as a contour integral as follows:
\begin{equation}\label{eq:area}
\int \D A = -\frac{1}{2}\oint \D s\,\bm{N}\cdot\bm{r}\;,	
\end{equation}
where the minus sign results from the convention of choosing the normal vector directed toward the interior of the curve.

\section{\label{sec:stm}Simple tension model}

\begin{figure}[t]
\centering
\includegraphics[width=0.7\textwidth]{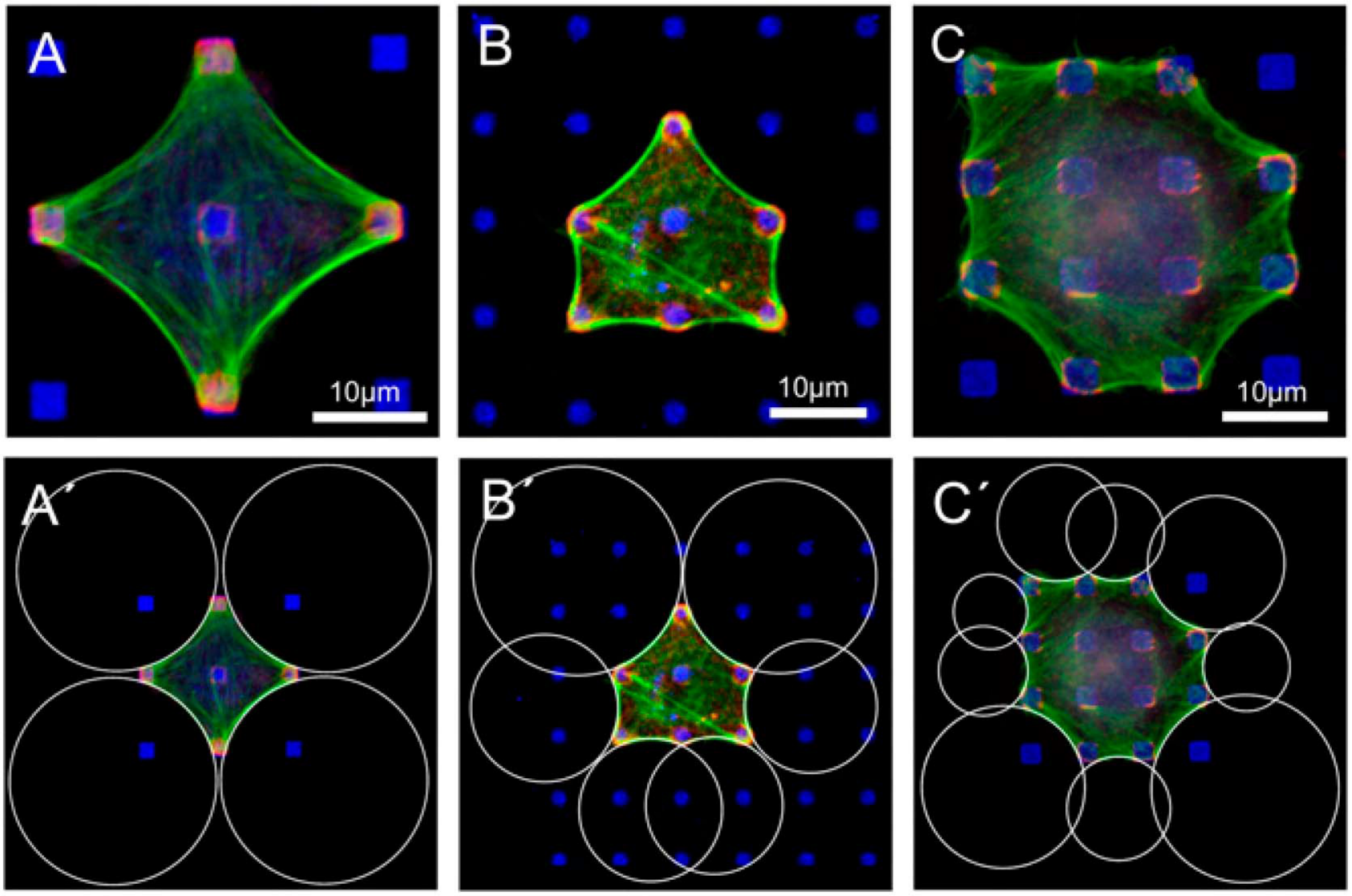}
\caption{\label{fig:bischofs}Cell shape on micropatterned substrates. (A-C) Arc-like contours composed of actin fibers characterize the shape of BRL (A and B) and B16 cells (C) cultured on substrates of micropatterned fibronectin dots. Cultures were labeled for actin (green), paxillin (red), and fibronectin (blue). Scale bars 10 $\upmu$m. (A'–C') For all cases, arc-like contours fit well to circles determined by custom-made software. (B and C) The circles spanning diagonal distances show larger radii than the circles spanning the shorter distances between neighboring adhesions. Reproduced from Ref. \cite{Bischofs:2008}.}
\end{figure}

At time scales when the cell is fully spread, the forces generated by the actomyosin cytoskeleton are mainly contractile and this gives rise to an {\em effective tension} that is transmitted to the substrate thorough the focal adhesions \cite{BarZiv:1999}. The actin cortex localized at the cell periphery naturally resists to this inward contraction by generating a competing contractile force, which, in turn, balances the bulk contractility leading to an equilibrium configuration. The overall effect of actomyosin contractily and adhesion on the shape of the cell was investigated by Bischofs {\em et al}. in Refs. \cite{Bischofs:2008,Bischofs:2009} by mean of a simple and yet very rich mechanical model known as {\em simple tension model}.

If the contractile forces acting in the interior of the cell are isotropic, the difference $\bm{\hat{\Sigma}}_{\rm out}-\bm{\hat{\Sigma}}_{\rm in}$ between the stresses across the cell edge can be reasonably approximated as an active pressure, namely:
\begin{equation}\label{eq:active_pressure}
\bm{\hat{\Sigma}}_{\rm out}-\bm{\hat{\Sigma}}_{\rm in}=\sigma\bm{\hat{I}}\;,
\end{equation}
where $\bm{\hat{I}}$ is the two-dimensional identity matrix. Similarly, the stress resultant produced by the contraction of peripheral actin can be expressed as $\bm{F}=\lambda\bm{T}$, where $\lambda=\lambda(s)$ is an effective interfacial tension. In the absence of external forces, Eqs. \eqref{eq:force_balance} and \eqref{eq:frenet} yield:
\begin{equation}
\lambda'\bm{T} + (\lambda+\kappa\sigma) \bm{N} = \bm{0}\;,
\end{equation}
along each individual cellular arc connecting two consecutive adhesion points. Because $\bm{T}$ and $\bm{N}$ are orthogonal, both terms must vanish. Thus, mechanical equilibrium requires $\lambda={\rm const}$ and:
\begin{equation}\label{eq:young_laplace}
\kappa = -\frac{\sigma}{\lambda}\;,
\end{equation}
where the negative sign reflects that the cell is everywhere concave, with exception for a discrete number of adhesion points. The same result could have also been obtained from a minimization of an effective energy functional of the from:
\begin{equation}\label{eq:energy}
E[\bm{r}] = \lambda \oint \D s + \sigma \int \D A\;.
\end{equation}
Indeed, upon performing a small normal variation of the curve, $\bm{r}\rightarrow\bm{r}+\epsilon\bm{N}$, standard manipulations of Eq. \eqref{eq:energy} yield:
\begin{equation}
\delta E = E[\bm{r}+\epsilon\bm{N}]-E[\bm{r}] = - \oint \D s\,(\sigma+\lambda\kappa)\epsilon\;.
\end{equation}
Thus Eq. \eqref{eq:young_laplace} identifies a minimizer of the energy functional Eq. \eqref{eq:energy}. Eq. \eqref{eq:force_balance} is, however, more generic and, as we will see in Sec. \ref{sec:atm}, allows to account for active stresses that could not be constructed from variational principles.

The quantities $\sigma$ and $\lambda$ are material parameters that embody the biomechanical activity of myosin motors in the actin cytoskeleton. Treating cells with pharmacological drugs able to disrupt the cytoskeleton allow some degree of manipulation of these quantities. Cells treated with Y-27632, a general inhibitor of the Rho-kinase pathway, and blebbistatin, a specific inhibitor of nonmuscle myosin II, have been reported to invaginate more than untreated cells, suggesting a strong reduction of $\lambda$ accompanied by the presence of a residual $\sigma$ \cite{Bischofs:2008}. The competition between bulk and peripheral contractility along the cell boundary results, therefore, in the formation of arcs of constant curvature, through a mechanism analogous to the Young-Laplace law for fluid interfaces. The shape of the cell boundary is then approximated by a sequence of circular arcs, whose radius $R=1/\kappa$ might or might not be uniform across the cell, depending on how the cortical tension $\lambda$ varies from arc to arc (Fig. \ref{fig:bischofs}).

\begin{figure}[t]
\centering
\includegraphics[width=\textwidth]{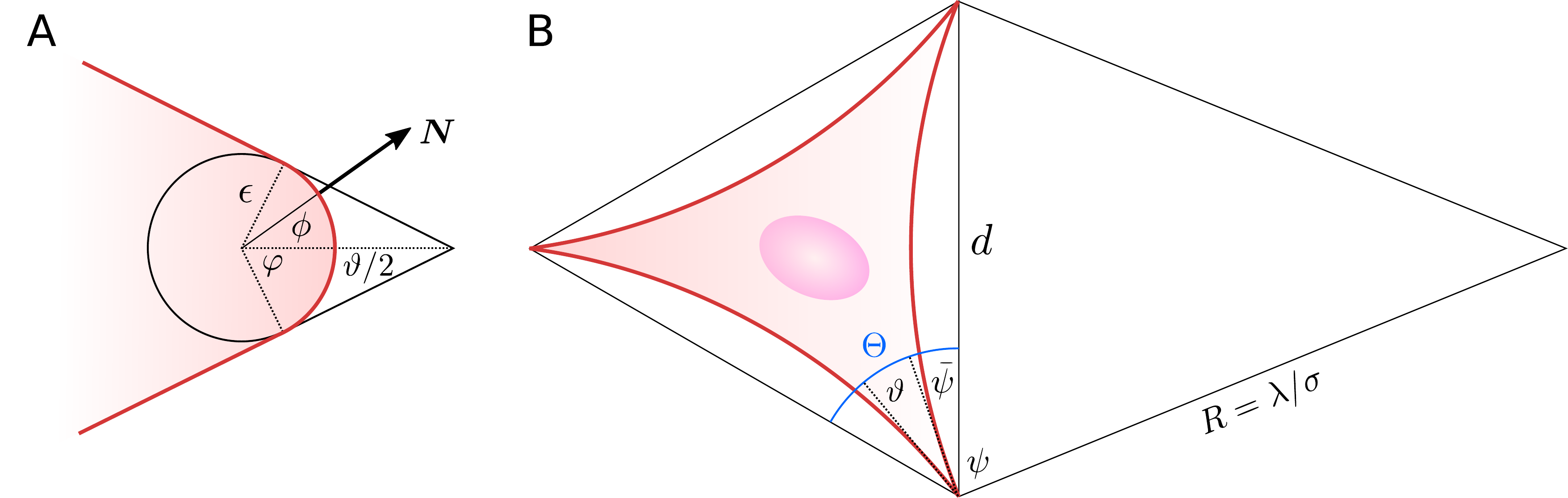}
\caption{\label{fig:traction}(A) To calculate the traction force exerted by the cell at a point of adhesion, it is convenient to approximate a kink with a circular arc of radius $\epsilon$ and take the limit of $\epsilon\rightarrow 0$. This procedure yields Eq. \eqref{eq:traction}. (B) Illustration of the calculation summarized by Eqs. \eqref{eq:trigonometric} and \eqref{eq:traction_vs_d}.}
\end{figure}

Although deliberately simple-minded, the picture outline so far is still sufficiently complex to account for some of the peculiar mechanical properties that are unique to living materials, such as the ability of regulating the forces exerted on a substrate depending on its stiffness. To appreciate this, let us consider again Eq. \eqref{eq:force_balance}. In between two adhesion points force balance results in the invagination of the cellular arcs as demanded by Eq. \eqref{eq:young_laplace}. At the adhesion points, on the other hand, both contractile forces exerted by the bulk and peripheral cytoskeleton are counterbalanced by the substrate, so that $\bm{f}_{\rm ext}(\bm{r})=-\sum_{i}\bm{F}_{\rm traction}(\bm{r})\delta(\bm{r}-\bm{r}_{i})$, where the sum runs over all adhesion points and the negative sign reflects that the force exerted by the cell on the substrate is equal and opposite to the force exerted by the substrate on the cell. Thus, from Eq. \eqref{eq:force_balance}, we have:
\begin{equation}\label{eq:local_traction}
\sum_{i}\bm{F}_{\rm traction}(\bm{r})\delta(\bm{r}-\bm{r}_{i})=(\sigma+\lambda\kappa)\bm{N}\;.
\end{equation}
Note that both sides of this equation diverge at the adhesion point as $\kappa\rightarrow\infty$ at a kink. In order to calculate the traction force exerted by a specific focal adhesion, one can approximate the kink as a circle of radius $\epsilon$, hence curvatures $\kappa=1/\epsilon$ (Fig. \ref{fig:traction}A). Then, taking $\bm{N}=(\cos\phi,\sin\phi)$, with $-\varphi \le \phi \le \varphi $, integrating Eq. \eqref{eq:local_traction} along the circle and taking the limit of $\epsilon\rightarrow 0$ yields \cite{Bischofs:2009}:
\begin{equation}\label{eq:traction}
\bm{F}_{\rm traction} 
= \lim_{\epsilon \rightarrow 0} \int_{-\varphi}^{\varphi} {\rm d\phi}\,\epsilon\,\left(\sigma+\frac{\lambda}{\epsilon}\right)\bm{N} 
= \bm{F}(\varphi)+\bm{F}(-\varphi)
= 2\lambda \cos\frac{\vartheta}{2}\,\bm{\hat{x}}\;,
\end{equation}
where $\vartheta=\pi-2\varphi$ is the opening angle of the kink (Fig. \ref{fig:traction}A). 

This simple considerations illustrate the essence of adaptive mechanical response in cells. According to Eq. \eqref{eq:traction}, the more acute is the kink (the smaller is $\vartheta$) the larger is the force exerted by the cell. The maximum traction $\bm{T}=2\lambda\bm{\hat{x}}$ is attained when $\vartheta=0$, thus the kink reduces to a cusp and all the tension exerted by the peripheral actin is employed to deform the substrate. Now, the opening angle $\vartheta$ at a given adhesion point, is not a free parameter, but is set by the curvature and the length of the cellular arcs that are directly connected to it. To see this let us consider a simple triangular cell whose adhesion points are located at a distance $d=2R\cos\psi$, with $R=\lambda/\sigma$ and $\psi$ as illustrated in Fig. \ref{fig:traction}B, from one another and let $\bar{\psi}$ be the internal angle of the triangle identified by the convex hull of the cell at a specific adhesion point (Fig. \ref{fig:traction}B). The opening angle is then $\vartheta=\Theta-2\bar{\psi}$, with $\psi+\bar{\psi}=\pi/2$. Therefore:
\begin{equation}\label{eq:trigonometric}
\cos\frac{\vartheta}{2} = \cos\frac{\Theta}{2}\sin\psi+\sin\frac{\Theta}{2}\cos\psi\;.
\end{equation}
Finally, taking $\cos\psi=d\sigma/(2\lambda)$, one can express the traction force as:
\begin{equation}\label{eq:traction_vs_d}
\bm{F}_{\rm traction} = 2\lambda\left[\left(\frac{d\sigma}{2\lambda}\right)\sin\frac{\Theta}{2}+\sqrt{1-\left(\frac{d\sigma}{2\lambda}\right)^{2}}\,\cos\frac{\Theta}{2}\right]\bm{\hat{x}}\;.
\end{equation}
Thus, for fixed $\sigma$ and $\lambda$ values, the more distant are the adhesion points, the more acute is the opening angle and the larger is the force exerted by the cell. Furthermore, if the substrate is compliant, the distance $d$ between focal adhesion depends on how much it deforms under the effect of traction forces. In particular, the softer the substrate the closer are the adhesion points, the weaker is the force exerted by the cell. Vice-versa, on stiffer substrates $d$ will be larger and the cell is expected to exert more force. These predictions, which are verified in experiments with fibroblasts and endothelial cells plated on continuous substrates of various rigidity \cite{Lo:2000,Yeung:2005}, provide a simple and yet insightful example of how in the interplay between cellular geometry and the active contraction provided by the actin cytoskeleton, can lead to an adaptive mechanical behavior even in the absence of biochemical regulation. 

\section{\label{sec:atm}Anisotropic tension model}

\begin{figure}[ht!]
\centering
\includegraphics[width=\textwidth]{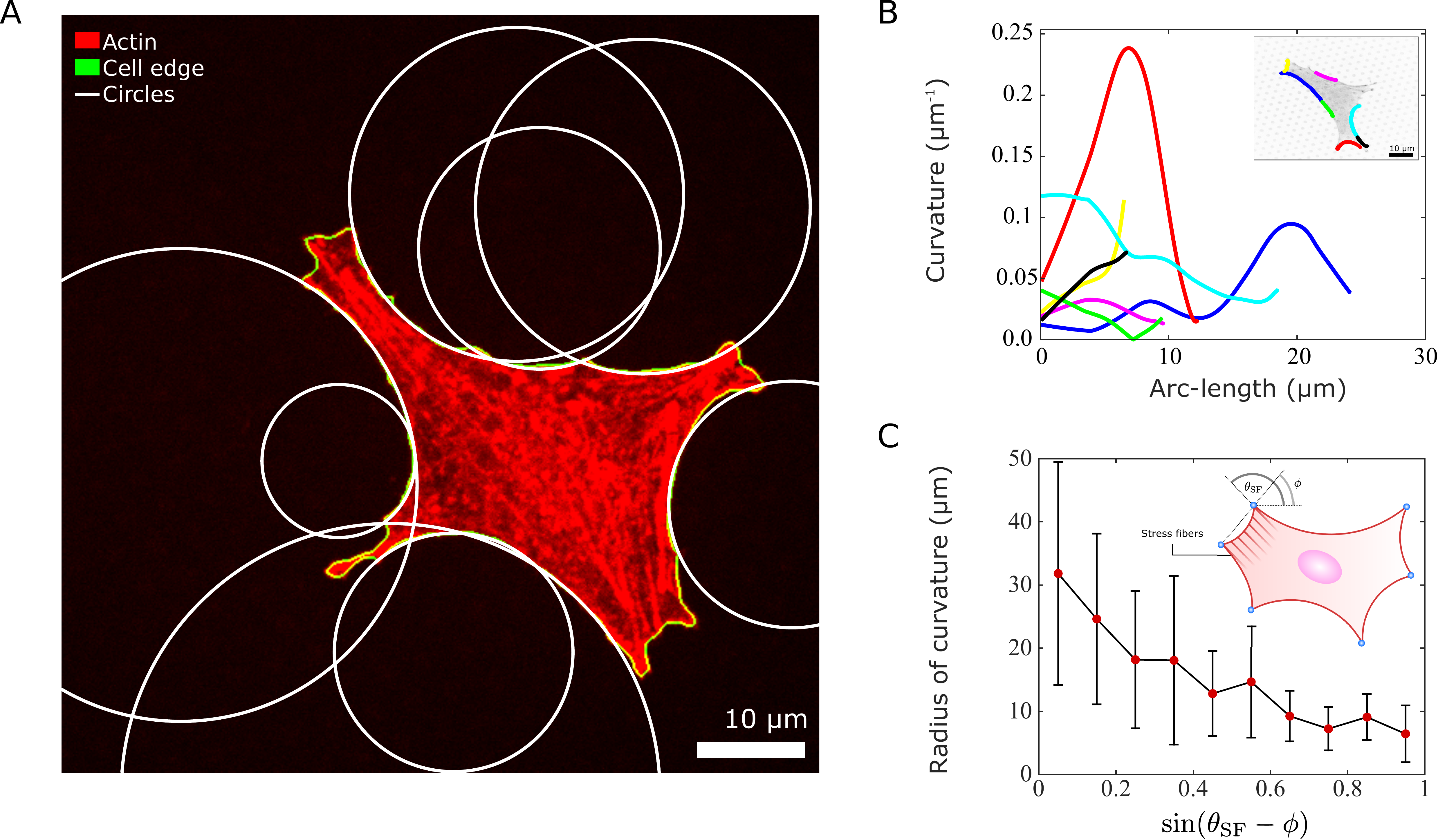}
\caption{\label{fig:anisotropic}Relation between stress fibers and curvature of the cell edge. (A) A cell with an anisotropic actin cytoskeleton (epithelioid GE$\upbeta$3) with circles (white) fitted to its edges (green). The end-points of the arcs are identified based on the forces exerted on the pillars. The actin cytoskeleton is visualized with TRITC-Phalloidin (red). Scalebar is 10 $\upmu$m. (B) Curvature versus arc-length for a specific cell (inset). Longer arcs, whose length is much larger than their average radius of curvature (i.e. $L \gg 1/|\kappa|$), are evidently non-circular as indicated by the fact that their curvature smoothly varies by a factor two along the arc. (C) Arc radius as a function of the sine of the angle $\theta_{\rm SF}-\phi$, between the local orientation of the stress fibers and that of the distance between the adhesion points (data show the mean $\pm$ standard deviation). Adapted from Ref. \cite{Pomp:2018}.}	
\end{figure}

Many cells, including the fibroblastoids (GD$\upbeta$1, GD$\upbeta$3) and epithelioids (GE$\upbeta$1, GE$\upbeta$3) displayed in Fig. \ref{fig:anisotropic}A \cite{Pomp:2018,Danen:2002}, develop directed forces by virtue of the strong anisotropic cytoskeleton originating from the actin stress fibers \cite{Burridge:2013,Pellegrin:2007}. This scenario is, evidently, beyond the scope of the simple tension model reviewed in Sec. \ref{sec:stm}. In these cells, the longer arcs appear indeed prominently non-circular, as indicated by the fact that their curvature smoothly varies along the arc by a factor two  (Fig. \ref{fig:anisotropic}B). On the other hand, the average radius of curvature of the cellular arcs appears significantly correlated with the orientation of the stress fibers. In particular, the radius of curvature decreases as the stress fibers become more perpendicular to the cell cortex (Fig. \ref{fig:anisotropic}C). This correlation is intuitive as the bulk contractile stress focusses in the direction of the stress fibers.

The anisotropy of the actin cytoskeleton can be incorporated into the contour models by modelling the stress fibers as contractile force-dipoles. As it is known from the literture on active fluids \cite{Pedley:1992,Simha:2002}, this collectively gives rise to a directed contractile bulk stress, such that
\begin{equation}\label{eq:active_stress}
\bm{\hat{\Sigma}}_{\rm out}-\bm{\hat{\Sigma}}_{\rm in} = \sigma\bm{\hat{I}}+\alpha\bm{n}\bm{n}\;,
\end{equation}
with $\alpha>0$ the magnitude of the directed contractile stress and $\bm{n}=(\cos\theta_{\rm SF}, \sin\theta_{\rm SF})$ the average direction of the stress fibers (Fig. \ref{fig:anisotropic}C inset). The ratio between isotropic contractility $\sigma$ and directed contractility $\alpha$ measures the degree of anisotropy of the bulk stresses. With this stress tensor the force balance equation \eqref{eq:force_balance} becomes:
\begin{equation}\label{eq:atm}
\lambda'\bm{T}+(\sigma+\lambda\kappa)\bm{N}+\alpha(\bm{n}\cdot\bm{N})\bm{n} = \bm{0}\;.
\end{equation}
Because $\bm{n}$ has, in general, non-vanishing projections on both the tangent and normal directions of the cell edge, this condition implies that  in the presence of an anisotropic cytoskeleton, the cortical tension $\lambda$ is no longer constant along the cell cortex. 

\begin{figure}[ht!]
\centering
\includegraphics[width=\textwidth]{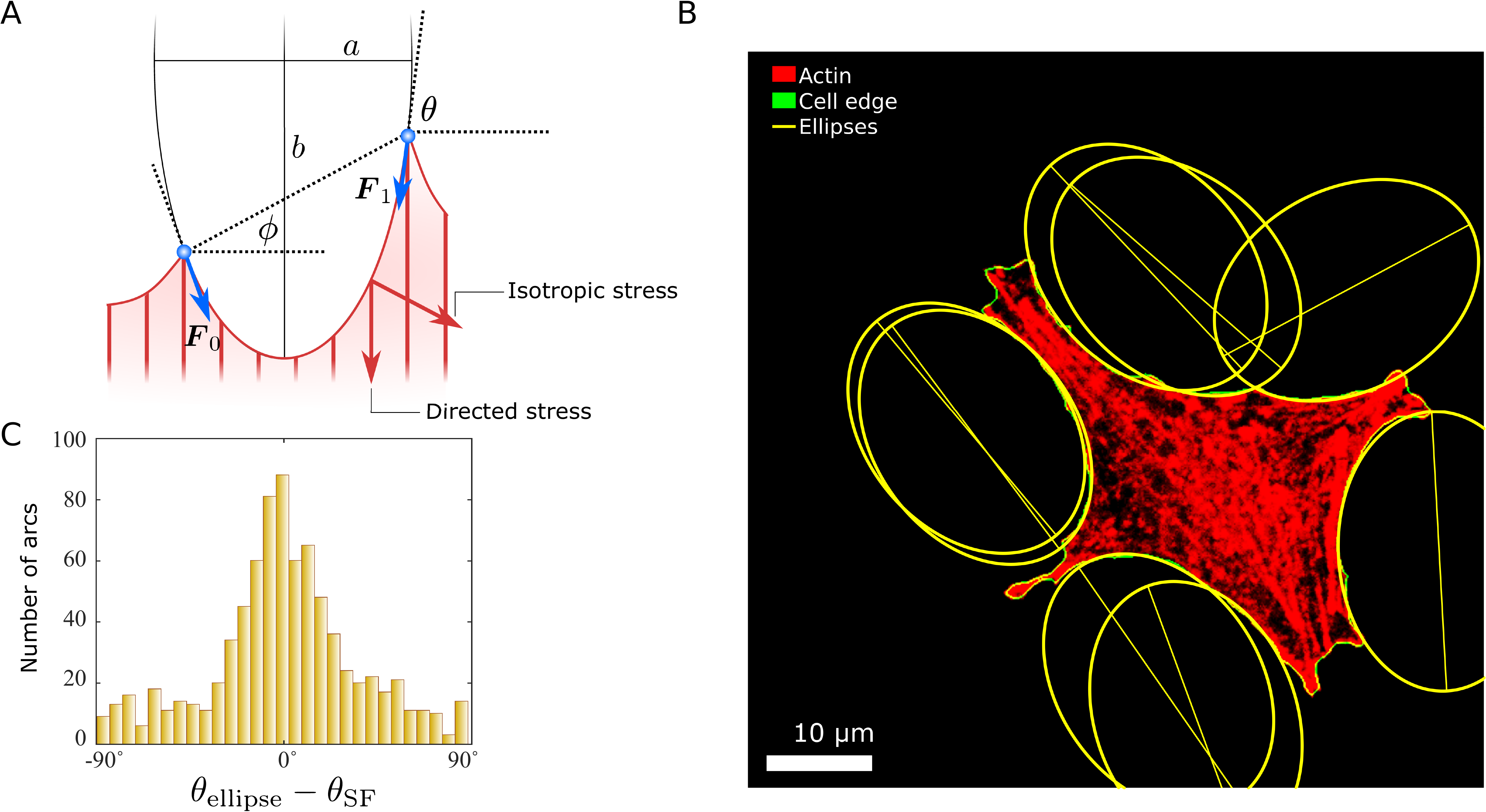}
\caption{\label{fig:ellipse}The anisotropic cytoskeleton is reflected in the elliptical shape of the cell edge. (A) Schematic representation of our model for $\theta_{\rm SF} = \pi/2$. A force balance between isotropic stress, directed stress and line tension results in the description of each cell edge segment (red curve) as part of an ellipse of aspect ratio $a/b=\sqrt{\gamma}$, unique to each cell. The cell exerts forces $\bm{F}_{0}$ and $\bm{F}_{1}$ on the adhesion sites (blue). (B) An epithelioid cell (same cell as in Fig. \ref{fig:anisotropic}) with a unique ellipse (yellow) fitted to its edges (green). The end-points of the arcs are identified based on the forces exerted on the pillars. The orientations of the major axes (yellow lines) are parallel to the local orientations of the stress fibers. Scalebar is 10 $\upmu$m. (C) Histogram of $\theta_{\rm ellipse}-\theta_{\rm SF}$, with $\theta_{\rm ellipse}$ the orientation of the major axis of the fitted ellipse and $\theta_{\rm SF}$ the measured orientation of the stress fibers. The mean of this distribution is $0^{\circ}$ and the standard deviation is $36^{\circ}$. Reproduced from Ref. \cite{Pomp:2018}.}	
\end{figure}

It is useful to introduce a number of simplifications, with the goal of highlighting the physical mechanisms entailed in Eq. \eqref{eq:atm}. As the orientation of the stress fibers varies only slightly along a single cellular arc, one can assume $\theta_{\rm SF}$ to be constant along each arc, but different, in general, from arc to arc. Moreover, as all the arcs share the same bulk, we consider the bulk stresses $\sigma$ and $\alpha$ uniform throughout the cell. Let us then look at a specific cellular arc and, without loss of generality, choose to orient the cell in such a way the stress fibers are parallel to the $y-$axis. Thus $\vartheta_{\rm SF}=\pi/2$ (Fig. \ref{fig:ellipse}A).  Using Eq. \eqref{eq:frenet} and taking advantage of the fact that $\bm{n}$ does not change along the arc, one can express all terms in Eq. \eqref{eq:atm} as total derivatives and integrate the equation directly. This yields:
\begin{equation}
\lambda\bm{T}+(\sigma\bm{\hat{I}}+\alpha\bm{n}\bm{n})\cdot\bm{r}^{\perp} = \bm{C}_{1}\;.	
\end{equation}
where $\bm{r}^{\perp}=(-y,x)$ and $\bm{C}_{1}=(C_{1x},C_{1y})$ is an integration constant. Then, using $\bm{n}=\bm{\hat{y}}$ and $\bm{T}=(\cos\theta,\sin\theta)$, we can simply this as:
\begin{subequations}\label{eq:scalar_force_balance}
\begin{align}
\lambda\cos\theta &= C_{x}+\sigma y\,\\[5pt]
\lambda\sin\theta &= C_{y}-(\alpha+\sigma)x\;,
\end{align}
\end{subequations}
from which, using $\tan\theta=\D y/\D x$ and integrating, we obtain a general solution of the force-balance equation in the form:
\begin{equation}\label{eq:ellipse}
\frac{x^{2}}{\gamma}+y^{2} - \frac{2C_{1y}}{\sigma}\,x+\frac{2C_{1x}}{\sigma}\,y = C_{2}\;,	
\end{equation}
where $\gamma=\sigma/(\alpha+\sigma)$ and $C_{2}$ is another integration constant. Notice that, if both $\sigma$ and $\alpha$ are positive for a contractile system, $\gamma<1$. Eq. \eqref{eq:ellipse} describes an ellipse whose minor and major semi-axes are $a=\sqrt{\gamma C_{2}}$ and $b=\sqrt{C_{2}}$ respectively and whose center is determined by $\bm{C}_{1}$. For simplicity, we can choose the origin of our reference frame to coincide with the center of the ellipse, so that $\bm{C}_{1}=\bm{0}$. Using again Eq. \eqref{eq:scalar_force_balance} with $\tan\theta=-x/(\gamma y)$, we can further obtain an expression for the cortical tension as a function of the turning angle $\theta$, namely:
\begin{equation}
\frac{\lambda^{2}}{\sigma^{2}} = C_{2}\,\frac{1+\tan^{2}\theta}{1+\gamma\tan^{2}\theta}\;.
\end{equation}
This expression highlights the physical meaning of the constant $C_{2}$. As the right-hand side attains its minimal values when $\theta=0$, thus when tangent vector is perpendicular to the stress fibers, $C_{2}$ is related with the minimal tension $\lambda_{\min}$ withstood by the cortical actin, namely $C_{2}=\lambda_{\min}^{2}/\sigma^{2}$, so that the shape of the cellular arc is described by the implicit equation:
\begin{equation}\label{eq:ellipse_final}
\frac{\sigma^{2}}{\gamma\lambda_{\min}^{2}}\,x^{2}+\frac{\sigma^{2}}{\lambda_{\min}^{2}}\,y^{2} = 1\;,	
\end{equation}
and the tension withstood by the cortical actin is given, as a function of the turning angle, by:
\begin{equation}\label{eq:tension}
\lambda = \lambda_{\min} \sqrt{\frac{1+\tan^{2}\theta}{1+\gamma\tan^{2}\theta}}\;.
\end{equation}
In summary, in the presence of directed stresses the equilibrium conformation of the cell edge consists of arcs of an ellipse of semi-axes $a=\sqrt{\gamma}\,\lambda_{\min}/\sigma$ and $b=\lambda_{\min}/\sigma$ and whose major axis is parallel to the stress fibers. The dimensionless quantity $\gamma$ highlights the interplay between the forces experienced by the cell edge and its shape: on the one hand, $\gamma$ characterizes the anisotropy of the bulk stress,  on the other hand it determines the anisotropy of the cell shape. 

The key prediction the anisotropic tension model is illustrated in Fig. \ref{fig:ellipse}B, where the contour of the same cell shown in Fig. \ref{fig:anisotropic}A has been fit with ellipses \cite{Pomp:2018}. Whereas large variations in the circles' radii were required in Fig. \ref{fig:anisotropic}A, a unique ellipse ($\gamma=0.52$, $\lambda_{\min}/\sigma=13.4\,\upmu{\rm m}$) faithfully describes all the arcs in the cell. Fig. \ref{fig:ellipse}C shows the distribution of the difference between the orientation $\theta_{\rm ellipse}$ of the major axis of the fitted ellipse and the measured orientation $\theta_{\rm SF}$ of the stress fibers. The distribution peaks at $0^{\circ}$ and has a width of $36^{\circ}$, demonstrating that the orientation of the ellipses is parallel, on average, to the local orientation of the stress fibers as predicted by the model. 

The traction forces $\bm F_{0}$ and $\bm F_{1}$ can be straightforwardly calculated from Eqs. \eqref{eq:atm} and \eqref{eq:tension} in the form:
\begin{subequations}
\begin{align}
\frac{\bm F_0}{\lambda_{\min}}&=\left(\beta\sin\phi+\zeta\cos\phi\right)\bm{\hat x}+\left(-\frac\beta\gamma\cos\phi+\zeta\sin\phi\right)\bm{\hat y}\;,\\
\frac{\bm F_1}{\lambda_{\min}}&=\left(\beta\sin\phi-\zeta\cos\phi\right)\bm{\hat x}+\left(-\frac\beta\gamma\cos\phi-\zeta\sin\phi\right)\bm{\hat y}\;,
\end{align}
\end{subequations}
where:
\begin{equation}
\beta=\frac{d}{2b}\;,\qquad\qquad
\zeta=\sqrt{\frac{1+\tan^2\phi}{1+\gamma\tan^2\phi}-\frac{\beta^2}{\gamma}}\;.
\end{equation}
Here $d$ is the distance between the positions of both forces on the ellipse, $b$ is the major semi-axis of the ellipse and $\phi$ is the angle that the line through both points makes with the $x-$axis (see Fig. \ref{fig:ellipse}A).

\section{\label{sec:bending}The effect of bending elasticity}

The contour models reviewed in Secs. \ref{sec:stm} and \ref{sec:atm} postulate that the stresses arising in the actin cortex are purely contractile, thus tangential to the cell edge. Non-tangential stresses can build up as a consequence of the bending elasticity of the actin cortex as well as the plasma membrane. The effect of bending elasticity was considered in Ref. \cite{Banerjee:2013} in order to account for the prominent polymorphism observed in experiments on cardiac myocytes adhering to substrates of varying stiffness \cite{Chopra:2011}. In this work, myocytes grown on substrates having material properties mimicking physiological stiffness ($5-10$ kPa), were observed to spread less and develop convex and well rounded morphologies. In contrast, while plated on stiffer gels or glass, the same cell type is more likely to exhibit a concave shape and greater spread area. This crossover from convex to concave, in particular, cannot be explained from the simple tension model or the anisotropic tension model discussed in the previous sections, as these lack of passive restoring forces able to contrast the formation of the highly curved regions (i.e. kinks) that characterize a closed plane curve whose curvature $\kappa$ is everywhere negative. In this respect, bending elasticity is the most natural choice among possible restoring mechanisms.

From a theoretical perspective, the problem arising by incorporating bending elasticity in a contour model of adherent cells, directly relates with another classic problem in mechanics: finding the shape of an infinitely long elastic pipe subject to uniform later pressure. This problem was formulated by Maurice L\'{e}vy in 1884 \cite{Levy:1884} and for over a century drew the attention of many researchers, due to its tremendous richness of polymorphic and multistable solutions \cite{Tadjbakhsh:1967,Flaherty:1972,Arreaga:2002,Vassilev:2008,Djondjorov:2011,Mora:2012,Giomi:2012,Giomi:2013}. Unlike the classic L\'{e}vy problem, however, the model proposed here for adhering cells does not involve any constraint on the length of the boundary, which is then only solely constrained by the adhesion with the substrate \cite{Giomi:2013}. This feature, introduces in the model a number of crucial mechanical properties, including an adaptive bending stiffness of the cell boundary.

\begin{figure}[t]
\centering
\includegraphics[width=\textwidth]{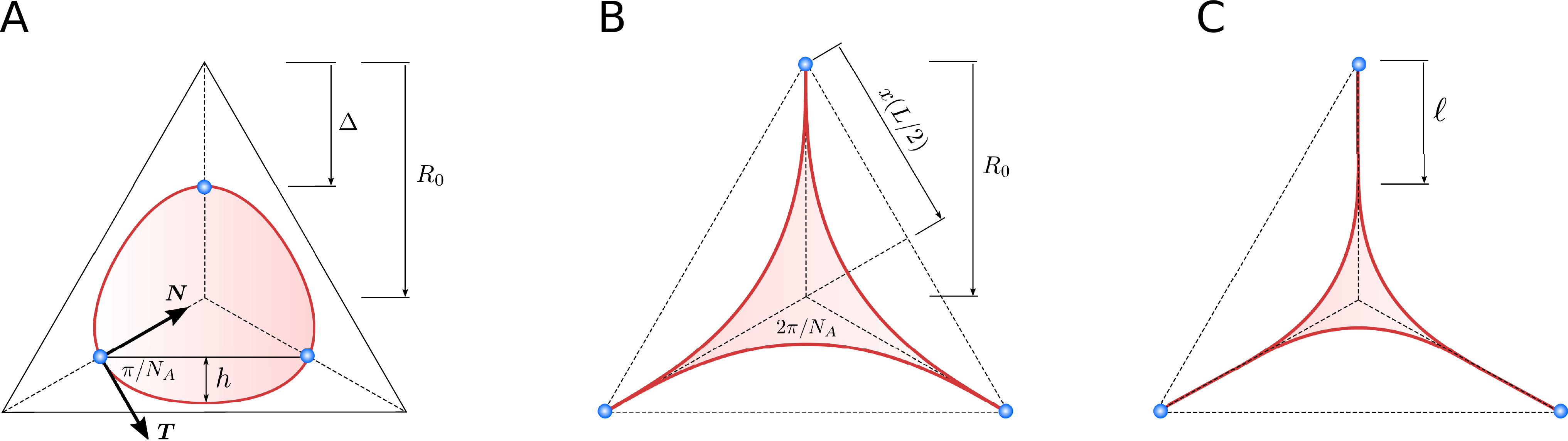}
\caption{\label{fig:bending_schematic}Cell anchored onto three pointwise adhesions located at the vertices of an equilateral triangle. (A) For small contractility values, the cell contour is everywhere convex with constant width. (B) When the contractility reaches a critical value $\sigma_{0}$, cell contour is purely concave with cusps at adhesion points. (C) For $\sigma>\sigma_{0}$, cusps gives rise to protrusion of length $\ell$. Adapted from Ref. \cite{Banerjee:2013}.}
\end{figure}

As it is known from classical elasticity of rods (see e.g. Ref. \cite{Landau:1970}), a slender structure forced to bend on the plane, is subject to a moment resultant
\begin{equation}
\bm{M} = B\kappa\bm{\hat{z}}\;.
\end{equation}
with $B$ the bending stiffness and $\kappa$ the curvature, with the usual sign convention introduced in Sec. \ref{sec:math}. Furthermore, the local balance of bending moments \cite{Landau:1970} requires:
\begin{equation}\label{eq:torque_balance}
\bm{M}'+\bm{T}\times\bm{F} = \bm{0}\;.
\end{equation}
Now, taking $\bm{F}=F_{T}\bm{T}+F_{N}\bm{N}$, with $F_{T}$ and $F_{N}$ the tangential and normal components of the stress resultant, and using $\bm{T}\times\bm{N}=\bm{\hat{z}}$, it one can cast Eqs. \eqref{eq:force_balance} and \eqref{eq:torque_balance} into:
\begin{subequations}\label{eq:tangent_normal_stress}
\begin{align}
B\kappa' + F_{N} &= 0\;,\\
F_{T}' - \kappa F_{N} + \bm{T}\cdot(\bm{\hat{\Sigma}}_{\rm out}-\bm{\hat{\Sigma}}_{\rm in})\cdot\bm{N} + \bm{T}\cdot\bm{f}_{\rm ext} &= 0\;, \\
\kappa F_{T} + F_{N}' + \bm{N}\cdot(\bm{\hat{\Sigma}}_{\rm out}-\bm{\hat{\Sigma}}_{\rm in})\cdot\bm{N}+\bm{N}\cdot\bm{f}_{\rm ext} &= 0\;. 
\end{align}
\end{subequations}
To make progress we restrict ourselves to cells with discrete rotational symmetry (Fig. \ref{fig:bending_schematic}A) and isotropic cytoskeleton. In this case one can assume $\bm{f}_{\rm ext}=f_{\rm ext}\bm{N}$. Then, Eqs. \eqref{eq:active_pressure}, (\ref{eq:tangent_normal_stress}a) and (\ref{eq:tangent_normal_stress}b) yield:
\begin{equation}
\bm{F} = \left(\lambda-\frac{1}{2}B\kappa^{2}\right)\bm{T}-B\kappa'\bm{N}\;,
\end{equation}
whereas Eq. (\ref{eq:tangent_normal_stress}c) yields an equation for the curvature $\kappa$:
\begin{equation}\label{eq:elastica}
B\left(\kappa''+\frac{1}{2}\kappa^{3}\right)-\lambda \kappa - \sigma - f_{\rm ext} = 0\;.	
\end{equation}
For $f_{\rm ext}={\rm const}$ this is the equation dictating the shape of an infinitely long pipe subject to a uniform pressure, or, alternatively, of a planar {\em elastica} spanned by a capillary film \cite{Mora:2012,Giomi:2012}. Unlike these examples, however, the length of the cell edge is not fixed and shall be determined from the balance between contractility, bending elasticity and the elastic response of the substrate embodied in the force per unit length $f_{\rm ext}$. 

To illustrate this last point, we can consider the simplified case in which peripheral contractility is negligible (i.e. $\lambda \approx 0$) and the cell periphery continuously adhere to the substrate along the cell edge. As there is no special direction on the plane, we can assume the cell to be a circle of radius $R$ centered at the origin. The force per unit length resulting form the deformation of the substrate, can then be expressed in the simple form:
\begin{equation}
f_{\rm ext} = - \frac{k_{s}(R-R_{0})}{\mathcal{L}}\;,
\end{equation}
where $k_{s}$ is the elastic stiffness of the substrate, $R_{0}$ is the radius of the cell before this starts stretching the substrate (i.e. once adhesions are formed and contractile forces start to build up) and $\mathcal{L}=2\pi R$ is the cell perimeter. Thus, setting $\kappa=1/R$ in Eq. \eqref{eq:elastica} yields the following cubic equation:
\begin{equation}\label{eq:radius}
(k_{s}+2\pi\sigma)R^{3}-k_{s}R_{0}R^{2}-\pi B = 0\;,
\end{equation}
The equation contains two length scales, $R_{0}$ and $\xi=(B/\sigma)^{1/3}$, and a dimensionless control parameter $k_{s}/\sigma$ expressing the relative amount of adhesion and contraction. For very soft anchoring $k_{s}\ll \sigma$ and Eq.~\eqref{eq:radius} admits the solution $R=\xi/2^{1/3}$. Thus non-adherent cells or cells adhering to extremely soft substrates (i.e. $k_s=0$), are predicted to have a radius of curvature that scales as $R \sim \sigma^{-1/3}$. The same scaling law is also predicted using \textit{active cable network} models of an adherent cell~\cite{Torres:2012}. If the cell is rigidly pinned at adhesion sites, $k_{s}\gg \sigma$ and $R\rightarrow R_0$. For intermediate values of $k_{s}/\sigma$ the optimal radius $R$ interpolates between $\xi$ and $R_{0}$ and is an increasing function of the substrate stiffness $k_{s}$, in case $\xi<R_{0}$, or a decreasing function if $\xi>R_{0}$. For $\xi=R_{0}$, the lower and upper bound coincide, and the solution is $R=R_0$. In particular, the case $R_0>\xi$ reproduces the experimentally observed trend that cell projected area increases with increasing substrate stiffness before reaching a plateau at higher stiffnesses~\cite{Yeung:2005,Chopra:2011,Engler:2004}. 
The asymptotic behavior and various limits of the solution are well captured by the interpolation formula:
\begin{equation}
R\approx\frac{k_{s}R_{0}+6\pi\sigma\,\xi}{k_{s}+6\pi\sigma}
\end{equation}
indicating that larger surface tension, hence larger cell contractility $\sigma$ leads to lesser spread area, consistent with the experimental observation that myosin-II activity retards the spreading of cells~\cite{Wakatsuki:2003}. Standard stability analysis of this solution under a small periodic perturbation in the cell radius shows that the circular shape is always stable for any values of the parameters $\sigma$, $k_{s}$ and $R_{0}$.

\begin{figure}[t]
\centering
\includegraphics[width=\textwidth]{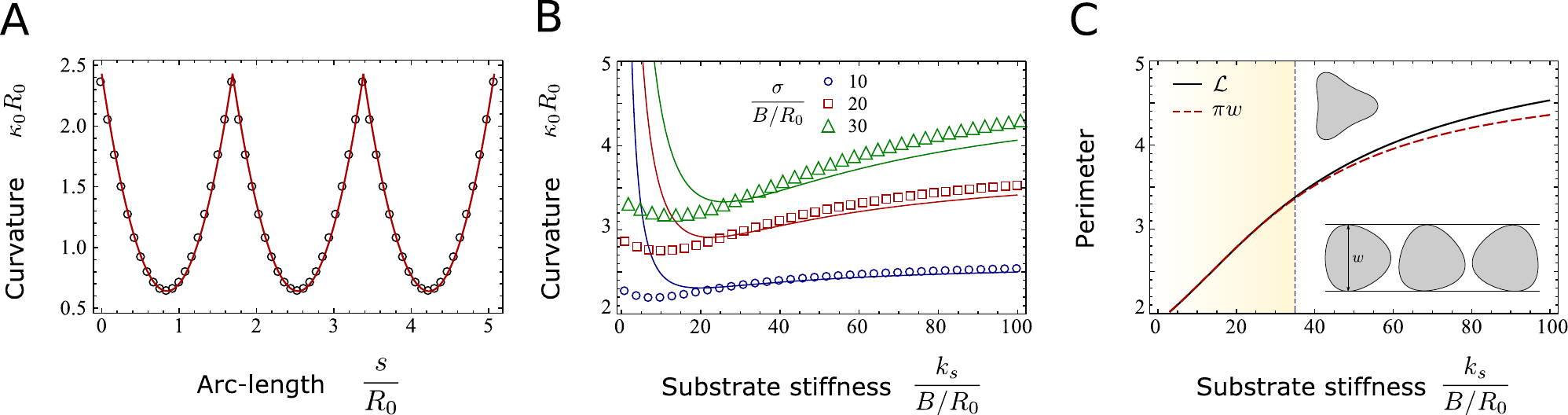}
\caption{\label{fig:curvature}Cell anchored onto three pointwise adhesions located at the vertices of an equilateral triangle. (A) Curvature versus arc-length for $\sigma R_{0}^{3}/B=10$, $k_{s}R_{0}^{3}=50$ and $N_{A}=3$. The circles are obtained from a numerical solution of Eq. \eqref{eq:elastica}, while the solid lines corresponds to our analytical approximation. (B) The end-point curvature $\kappa_{0}$ at the adhesion points as a function of the substrate stiffness for various contractility values. The points are obtained from numerical simulations while the solid lines correspond to our analytical approximation. (C) The total cell length $\mathcal{L}$ as a function of adhesion stiffness. For small stiffnesses the cell boundary form a curve of constant width (lower inset) and $\mathcal{L}=\pi w$, with $w$ the width of the curve. This property breaks down for larger stiffnesses when inflection points develops (upper inset). Reproduced from Ref. \cite{Banerjee:2013}.}
\end{figure}

For cells adhering to discrete number of adhesion sites, one can show that the circular solution for the cell boundary is never stable. For simplicity, we assume that $N_A$ adhesion sites are located at the vertices of a regular polygon of circumradius $R_{0}$ (Fig. \ref{fig:bending_schematic}A). The force per unit length exerted by the substrate is then:
\begin{equation}
f_{\rm ext} = -k_{s}\sum_{i=0}^{N_{A}-1}\delta(s-iL)[r(s)-R_{0}]\;,
\end{equation}
where $L$ the distance between consecutive adhesion points and $r(s)=|\bm{r}(s)|$. Using again the rotational symmetry of the problem, we can assume that the substrate is stretched at all adhesion points by the same amount. Then $r(s)-R_{0}=\Delta = {\rm const}$ and Eq. \eqref{eq:elastica} reduces to:
\begin{equation}\label{eq:euler-lagrange}
B\left(\kappa''+\frac{1}{2}\kappa^{3}\right)-\lambda \kappa - \sigma + k_{s}\Delta\sum_{i=0}^{N_{A}-1}\delta(s-iL) = 0\;.	
\end{equation}
Integrating Eq. \eqref{eq:euler-lagrange} along an infinitesimal neighborhood of the $i-$th adhesion point, one finds the following condition for the derivative of the curvature at the adhesion points:
\begin{equation}\label{eq:discontinuity}
\kappa'_{i}=-\frac{k_{s}}{2B}\,\Delta\;.
\end{equation}
The local curvature of the segment lying between adhesion points is, on the other hand, determined by the equation:
\begin{equation}\label{eq:levy}
\kappa''+\frac{1}{2}\kappa^{3}-\frac{\lambda}{B}\,\kappa-\frac{\sigma}{B}=0\;,
\end{equation}
with the boundary conditions : $\kappa(iL)=\kappa((i+1)L)=\kappa_0$, with $i=1,\,2\ldots\,N_{A}$ and $\kappa_{0}$ a constant to be determined. Without loss of generality we consider a segment located in the interval $0 \le s \le L$. Although an exact analytic solution of this nonlinear equation is available (e.g. Ref. \cite{Banerjee:2013}), an excellent approximation can be obtained by neglecting the cubic nonlinearity (Fig.~\ref{fig:curvature}A). With this simplification, Eq. \eqref{eq:euler-lagrange} admits a simple quadratic solution of the form:
\begin{equation}\label{eq:curvature}
\kappa(s)=\kappa_0 + \frac{\sigma}{2B}\,s(s-L)\;.
\end{equation}
Eqs. \eqref{eq:curvature} and \eqref{eq:discontinuity} immediately allow us to derive a condition on the cell perimeter, namely
\begin{equation}
L= \frac{k_{s}\Delta}{\sigma}\;.
\end{equation}
This leads, furthermore, to a linear relation between traction force $F_{\rm traction}=k_s \Delta$, and cell size :
\begin{equation}
F_{\rm traction}=\sigma L\;,
\end{equation}
which is indeed observed in traction force measurements on large epithelial cells~\cite{Mertz:2012}. 

To determine the end-point curvature $\kappa_{0}$, one can use the turning tangents theorem for a simple closed curve, Eq. \eqref{eq:turning_tangents}, which requires $\int_{0}^{L}\D s\,\kappa = 2\pi/N_{A}$. This leads to following relation between local curvature and segment length, or equivalently traction force, at the adhesion sites :
\begin{equation}
\kappa_{0}
=\frac{\sigma L^{2}}{12B} + \frac{2\pi}{N_{A}L}\;.
\end{equation}
A plot of $\kappa_{0}$ as a function of the substrate stiffness is shown in Fig. \ref{fig:curvature}B. Finally, to determine the optimal length of the cell segment $L$, we are going to make use of a remarkable geometrical property of the curve obtained from the solution of Eq.~\eqref{eq:euler-lagrange} with discrete adhesions: the fact of being a {\em curve of constant width} \cite{Gray:1997}. The width of a curve is the distance between the uppermost and lowermost points on the curve (see lower inset of Fig. \ref{fig:curvature}C). In general, such a distance depends on how the curve is oriented. There is however a special class of curves, where the width is the same regardless of their orientation. The simplest example of a curve of constant width is clearly a circle, in which case the width coincides with the diameter. A fundamental property of curves of constant width is given by the Barbier's theorem \cite{Gray:1997}, which states that the perimeter $\mathcal{L}$ of any curve of constant width is equal to width $w$ multiplied by $\pi$: $\mathcal{L}=\pi w$. As illustrated in Fig.~\ref{fig:curvature}C, this is confirmed by numerical simulations for low to intermediate values for contractility and stiffness. With our setting, the cell width is given by:
\begin{equation}\label{eq:width}
w = (R_{0}-\Delta)\left(1+\cos\frac{\pi}{N_{A}}\right)+h\left(\frac{L}{2}\right)\;,
\end{equation}
where $h(s)=\int_{0}^{s}\D s'\,\sin\theta(s')$ is the height of the curve above a straight line between two adhesions points (Fig. \ref{fig:curvature}A) and
\begin{equation}
\theta(s)=\int_{0}^{s}\D s'\,\kappa(s')=\theta_{0}+\kappa_{0}s+\frac{\sigma}{12B}\,s^{2}(2s-3L)
\end{equation}
the usual turning angle (Fig. \ref{fig:schematic}B). For small angles $h$ can be approximated as : 
\begin{equation}
h(s) \approx s(L-s)\left[\frac{\pi}{N_{A}L}-\frac{\sigma}{12B}\,s(L-s)\right]
\end{equation}
Using this, Eq. \eqref{eq:width} and the Barbier's theorem with $\mathcal{L}=N_{A}L$ allow us to obtain a quartic equation for the cell length:
\begin{equation}\label{eq:quartic}
\frac{N_{A}L}{\pi} = \left(1+\cos\frac{\pi}{N_{A}}\right)\left(R_{0}-\frac{\sigma L}{2k_{s}}\right) + \frac{L}{4}\left(\frac{\pi}{N_{A}}-\frac{1}{96}\frac{\sigma L^{3}}{B}\right)\;.
\end{equation}
Fig. \ref{fig:traction} shows plots of the traction $F_{\rm traction}=\sigma L$ with $L$ determined by solving Eq. \eqref{eq:quartic}. These results support the experimental trend that traction force increases monotonically with substrate stiffness $k_{s}$ before plateauing to a finite value for higher stiffnesses~\cite{Ghibaudo:2008,Mitrossilis:2009}. The plateau value increases with increasing contractility (Fig.~\ref{fig:traction}a). Traction force grows linearly with increasing contractility for $\sigma R_0^3/B \ll 1$, before saturating to the value $k_s R_0$ at large contractility $\sigma R_0^3/B \gg 1$, as shown in Fig.~\ref{fig:traction}b. Eq.~\eqref{eq:traction} is also consistent with experimentally observed trend that reducing contractility by increasing the dosage of myosin inhibitor Blebbistatin, leads to monotonic drop in traction forces~\cite{Mitrossilis:2009}.

\begin{figure}[t]
\centering
\includegraphics[width=\textwidth]{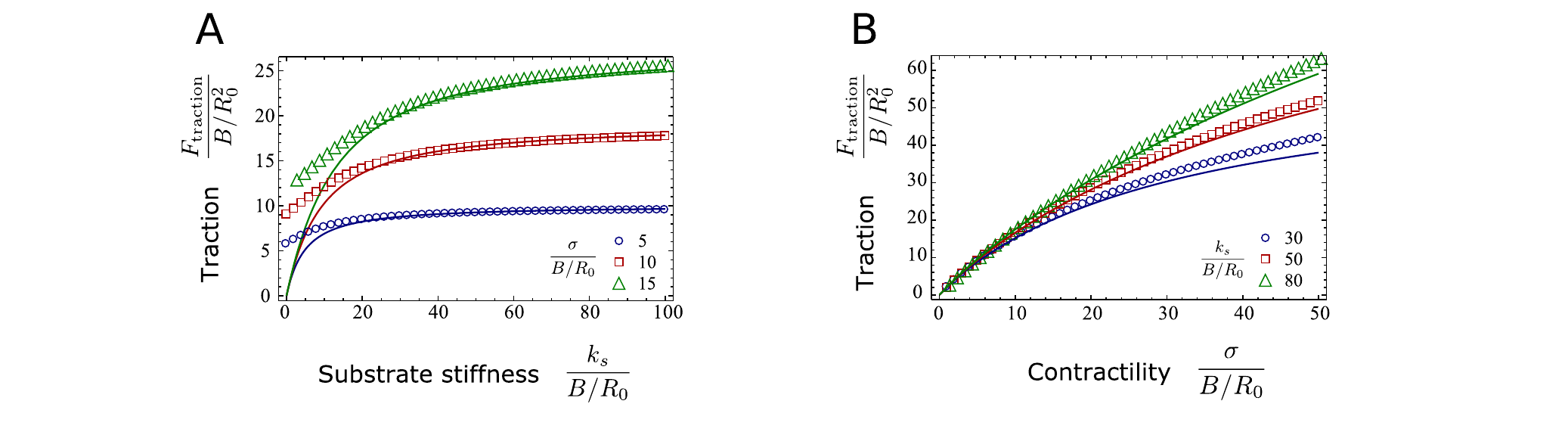}
\caption{\label{fig:traction}Traction force as a function of substrate stiffness (A) and contractility (B) obtained from a numerical solution of Eq. \eqref{eq:euler-lagrange}. Solid curves denote the approximate traction values obtained from Eq. \eqref{eq:quartic}. Adapted from Ref. \cite{Banerjee:2013}.}
\end{figure}


\begin{figure}[t]
\centering
\includegraphics[width=0.5\textwidth]{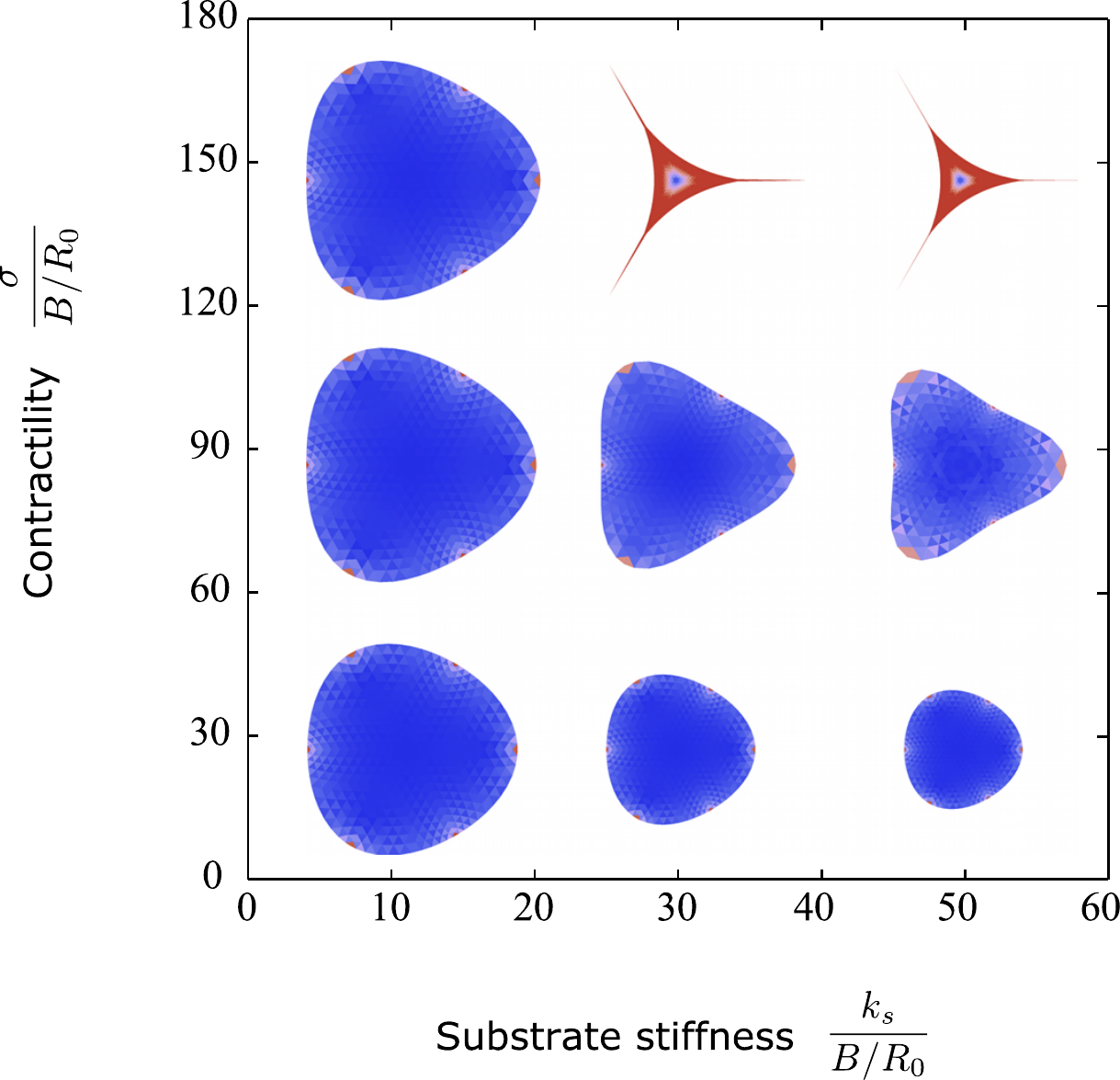}
\caption{\label{fig:phase_diagram}Phase diagram in $\sigma$-$k_s$ plane showing optimal configuration obtained by numerical minimization of the energy \eqref{eq:energy} for $N_A=3$. Adapted from Ref. \cite{Banerjee:2013}.}
\end{figure}

For low to intermediate values of $\sigma$ and $k_s$, cell shape is convex and has constant width. Upon increasing $\sigma$ above a $k_{s}-$dependent threshold, however, the cell boundary becomes inflected (see Fig. \ref{fig:phase_diagram} and upper inset of Fig. \ref{fig:curvature}C). Initially, a region of negative curvature develops in proximity of the mid point between two adhesions, but as the surface tension is further increased, the size of this region grows until positive curvature is preserved only in a small neighborhood of the adhesion points. Due to the presence of local concavities, the cell boundary is no longer a curve of constant width. 

If $\sigma$ is further increased, the inflected shape collapses giving rise to the star-shaped configurations shown in upper right corner of Fig. \ref{fig:phase_diagram} . These purely concave configurations are made by arcs whose ends meet in a cusp. The cusp is then connected to the substrate by a protrusion consisting of a straight segment strecthing until the adhesion point rest position, so that $\Delta = 0$ (Fig. ~\ref{fig:bending_schematic}C) and, consistent with Eq. \eqref{eq:discontinuity}, $\kappa={\rm const} = 0$ at the adhesion point. Unlike the previous transition from convex to non-convex shapes, this second transition occurs discontinuously and is accompanied by a region of bistability (see Ref. \cite{Banerjee:2013} for further detail). Away from the protrusion, the curvature has still the form given in Eq. \eqref{eq:curvature}, with $\kappa_{0}=0$ so that the boundary is everywhere concave or flat and the bending moment $\bm{M}=B\kappa\bm{\hat{z}}$ does not experience any unphysical discontinuity at the protrusions origin.

The length of the protrusion can be readily obtained using a similarity transformation. This construction, first investigated by Flaherty {\em et al.} \cite{Flaherty:1972} for the original L\'{e}vy problem, relies on the invariance of Eq. \eqref{eq:levy} under the following scaling transformation:
\begin{equation}\label{eq:scaling}
\{s,\kappa,\lambda,\sigma\} \rightarrow \left\{\Lambda s,\frac{\kappa}{\Lambda},\frac{\lambda}{\Lambda^{2}},\frac{\sigma}{\Lambda^{3}}\right\}\;.
\end{equation}
with $\Lambda$ a scaling factor. Calling then $\sigma_{0}$ the value of $\sigma$ at which the protrusion have zero length, the shape of the cell edge at any $\sigma>\sigma_{0}$ can be constructed starting from the reference shape illustrated in Fig. \ref{fig:bending_schematic}B as follows. First one calculates the scaling factor $\Lambda=(\sigma_{0}/\sigma)^{1/3}$ associated with the new $\sigma$ value. The reference shape is then rescaled by $\Lambda$ in such a way that the cusps are now disconnected from the original adhesion points. Finally, the cusps and the adhesion points are reconnected by straight segments of length $\ell_{\rm p}=R_{0}(1-\Lambda)$ (since $R_{0}$ is the circumradius of the reference shape and $\Lambda R_{0}$ that of the rescaled shape). This latter step, ultimately allows us to formulate a scaling law for the length of protrusions, namely:
\begin{equation}\label{eq:protrusion}
\frac{\ell}{R_{0}} = 1-\left(\frac{\sigma_{0}}{\sigma}\right)^{\frac{1}{3}}\;.	
\end{equation}
This transition from a smooth shape to a self-contacting shape with cusps is reminiscent of the post-buckling scenario of an elastic ring subject to a uniform pressure, but unlike this case, where the system undergoes a continuous transition from a simple curve to a curve with lines of contact, here the transition is discontinuous along both the loading branch (increasing $k_{s}$) and the unloading branch (decreasing $k_{s}$). The transition has moreover a strong topological character since it involves a jump in the rotational index of the curve, whose total curvature after the transition becomes:
\[
\oint_{\partial M}ds\,\kappa = \pi(2-N_{A})\;,
\]
by virtue of Eq. \eqref{eq:kinks}. Some further detail about the geometry of protrusions in this model can be found in Ref. \cite{Banerjee:2013}. 

\section{\label{sec:conclusions}Concluding remakrs}

In this chapter we have reviewed a simple theoretical framework for modeling the mechanical aspects of cell-substrate interaction. Several experimental works have demonstrated the latter to play a critical role in regulating a variety of cellular processes, from morphogenesis, motility, to cell linage and fate. While adhering to the extracellular matrix, cell develop specific morphologies depending on the geometrical and mechanical properties of their micro-environment. In turn, the forces arising inside the cell in response to this structural reorganization, drive biochemical cascades that not only feedback on the mechanical cell-matrix interaction, but also influence other processes such as cell cycle control and differentiation. Whereas the extraordinary complexity of these biomechanical pathways is still elusive, much progress has been made in understanding how the presence of actively generated contractile forces, on the one hand, and the absence of hard geometric constraints, on the other, give rise to a whole new class of mechanical phenomena commonly found in living systems, such as adaptivity, polymorphism, multistability etc. The simple contour models reviewed in this chapter, have given an important contribution in this respect, as they allow to cast the problem in a form that is often analytically tractable, thanks to the reduced dimensionality. 

Whereas insightful under many respects, contour models do not give access to the dynamics of the molecular process involved in cell adhesion and migration, such as the association dissociation events in the adhesion clusters or the regulation of myosin expression in the presence of various mechanical cues. This create the demand for a more comprehensive theoretical framework, where the continuum mechanics standpoint of contour models could be integrated into a multi-scale approach able to account for the mechanical, biochemical and genetic aspects of cellular organization alike.%
\\[10pt]
{\em Acknowledgements} I am indebted with Koen Schakenraad, Thomas Schmidt, Wim Pomp and Shiladitya Banerjee for contributing to the work reviewed here. This work is partially supported by the Netherlands Organisation for Scientific Research (NWO/OCW), as part of the Frontiers of Nanoscience program and the Vidi scheme.

\end{document}